\documentclass[sn-aps,pdflatex]{sn-jnl}

\usepackage{amsmath, amsthm, amssymb, amsfonts}
\usepackage{graphicx}
\usepackage{manyfoot}
\usepackage{xcolor}

  \newcommand{\revision}[1]{#1}

\begin{document}


\title[Stepped acoustic diode]{One-dimensional stepped chain of beads as a \revision{broadband} acoustic diode }

\author*[1,2]{\fnm{Carlos} \sur{Vasconcellos}}\email{carlos.vasconcellos@usach.cl}

\author[3]{\fnm{Stéphane} \sur{Job}}\email{stephane.job@isae-supmeca.fr}

\author[1,2]{\fnm{Francisco} \sur{Melo}}\email{francisco.melo@usach.cl}

\affil[1]{\orgname{Universidad de Santiago de Chile}, \orgdiv{Department of Physics}, \orgaddress{\street{Avenida \revision{V{\'i}ctor Jara} 3493}, \postcode{9170124}, \city{ Santiago}, \country{Chile}}}

\affil[2]{\orgname{Center for Soft Matter Research}, \orgdiv{SMAT-C}, \orgaddress{\street{Avenida Bernardo O'Higgins 3363}, \postcode{9170124}, \city{Estaci\'on Central}, \state{Santiago}, \country{Chile}}}

\affil[3]{\orgdiv{\orgname{ISAE-Supméca}, Quartz Lab (EA-7393)}, \orgaddress{\street{3 rue Fernand Hainaut} \postcode{93400} \city{Saint-Ouen-sur-Seine}, \country{France}}}


\abstract{Asymmetric transmission of waves has received considerable attention in the last few decades, inspired by numerous applications in the domains of optics and wave communications. \revision{A possible way to achieve such asymmetry in wave propagation is through the violation of reciprocity}. We explore asymmetric propagation in an acoustic diode (AD) made of two segments of equal sphere sizes but different elastic properties, in which both chains obey Hertzian's contact law. In the forward configuration, the first chain acts as the nonlinear medium, ensuring the suitable generation of a self-demodulated signal, while the second chain, made of soft spheres, provides the low-pass filter. It is shown that both large nonlinearity and low dispersion cause the smallest distortion of the wave envelope, therefore optimizing the transmission of information through the system.}


\maketitle


\section{Introduction}

Metamaterials are a heterogeneous group of artificial materials with special electromagnetic or mechanical properties that arise from the designed structure and not from its composition~\cite{Engheta2006}. More recently, metamaterials with new properties in the domain of acoustics have been designed~\cite{Fok2008}, leading to timbre-preserved non-reciprocal transmission capabilities of sound originating from musical instruments~\cite{Guo2023}. In the same vein, phononic crystals and acoustic metamaterials can be designed with a hierarchy of spectral characteristics, leading to a variety of functionalities, including frequency filtering, wave guiding, wavelength multiplexing, and demultiplexing~\cite{Deymier2013}. Metamaterials have been designed using broad strategies, currently based on the control of their physical structure. Thus, most metamaterials are often periodic at some suitable scale adapted to the range of the wavelength of application. Early examples include properties such as negative permittivity and permeability~\cite{Smith2000}, and a negative refraction index~\cite{Shelby2001,Smith2004}.

Reciprocity is the physical property that expresses the symmetrical transmission of waves between two points in space, thus being the result of the time-reversal symmetry of physical laws. However, in some applications, asymmetrical propagation is desirable instead. For instance, if the protection of a region of space or a selective filter is needed, it can be achieved by allowing the wave to propagate in only one direction~\cite{Fleury2014}. The design of metamaterials and strategies to produce asymmetric propagation operating for mechanical waves has been motivated by numerous applications in optics and radio waves~\cite{Engheta2006,Wu2018}.
For elastic waves, giant, broadband and robust non-reciprocity has been achieved for instance in nonlinear metamaterial, owing to specially designed asymmetric cell topology leading to amplitude-dependent band structure~\cite{Fang2020}.
Several mechanisms have been proposed to achieve asymmetric propagation, most of them based on the concept of a diode~\cite{Boechler2011,Liang2010,Popa2014,Fleury2014}, \revision{sometimes referred as a rectifier or as a switch}. The simplest sonic or acoustic diode (AD) is made of two components, namely, a strong nonlinear medium (NLM) and a composite device exhibiting a band structure; for instance, a sonic crystal (SC). The nonlinear medium is the key component responsible for symmetry breaking and the violation of the reciprocity theorem, which applies to linear wave propagation. While the sonic crystal serves as a selective filter, preventing fundamental frequency propagation and allowing the passage of the second harmonic wave (SHW). If both components are organized in a sequential configuration when the mechanical wave impinges first on the nonlinear medium, only the nonlinearly generated SHW can pass through the diode. This is achieved if the transmission properties of the sonic crystal are selected such that the SHW frequency is located in the allowed band, and the fundamental is in the band gap (BG) of the sonic crystal. However, a mechanical wave at the fundamental frequency, impinging the system in the opposite direction, is blocked, which prevents it from reaching the NLM. The described mechanism was first implemented by Liang et al.~\cite{Liang2010} in a 1D phononic crystal composed of alternating layers of glass and water coupled to a nonlinear acoustic medium that was prepared by mixing water and gas bubbles. However, acoustic rectification is obtained if the main-wave frequency is in the stop band of the superlattice and the SHW is located in the pass band. This is an important difference between an AD and its electrical or thermal counterpart, in which the whole signal is transmitted in the forward direction. In addition, in an AD, if the incident wave is coding some information, in the general case, this information is lost due to the nonlinear mode conversion and filtering.

To our knowledge, strategies for modulating the incident wave and its demodulation have not been explored yet. In addition to non-linearity, the diffraction effect has been explored to achieve optimized asymmetric propagation~\cite{Li2011} in a linear acoustic diode made of a 2D phononic crystal composed of square steel rods organized in two sections: a phononic crystal (a square lattice with a smooth and regular exterior surface) and a diffracting structure (a square lattice with an irregular, rough exterior surface providing diffracting sites). Waves with frequencies within the phononic band gap are not transmitted if they impinge on the phononic crystal (reverse direction), while they propagate if they impinge on the diffraction structure (direct direction). The operation of this diode is based on the fact that waves impinging on the smooth surface (the reverse direction) with frequencies within the forbidden band are reflected backward. On the contrary, waves impinging on the rough surface (the forward direction) are allowed to pass partially, due to sonic energy conversion and high-order diffractions, to modes with spatial frequencies capable of overcoming the barrier of the band gap. Remarkably, the incident waves do not alter their frequencies when transmitted through the structure.

Until the present, in addition to the mechanism introduced first by Liang~\cite{Liang2010}, two more related ideas for implementing an AD have been discussed~\cite{Maldovan2013}. In the work of Daraio et al.~\cite{Boechler2011}, the rectifier is made of a compressed one-dimensional array of spheres in contact, containing a lighter mass defect located close to a boundary. In the forward configuration, which corresponds to an input at the intruder side, if the device is excited at frequencies above the gap but near the resonance frequency of the defect, harmonics resulting from the combination of these frequencies can propagate through the array. In contrast, in the reverse configuration, with the defect being too far from the source, the array acts as a filter. Given their common working principle, these devices act as diodes only in a restricted band of frequencies. In addition, in the forward direction the carrier signal is also suppressed, and only the nonlinearly generated low-frequency component is allowed to pass through. This contrasts with the work of Liang et al.~\cite{Liang2010}, where only the SHW is transmitted. The question of how to achieve the coding of desired information in this latter mode seems difficult to address due to the complex combination of low-frequency modes. Devaux et al.~\cite{Devaux2015} proposed to achieve the conversion effect through a nonlinear self-demodulation mechanism in a 3D unconsolidated granular medium, while a phononic crystal acted as a selective filter. This design has two main advantages: first, there is no need to precisely tune the acoustic filter due to the frequency down-conversion of the self-demodulation effect; and second, in the forward configuration, the transmission of the self-demodulated wave instead of the SHW opens the possibility to code desired information through low-frequency amplitude modulation of the incident carrier wave.
Non-reciprocity is also at the heart of the so-called targeted energy transfer (TET) in nonlinear energy sinks (NES)~\cite{Vakakis2009}, which recently brought emerging researches in elastodynamics~\cite{Vakakis2022}. For instance, extreme break of reciprocity with large transmissibility in the prefered direction of propagation has been realized via the nonlinearity of a single element at the contact between two non-dispersive waveguides~\cite{Grinberg2018}, and via the use of a combination of nonlinear asymmetric gates~\cite{Michaloliakos2023}.  

\revision{More recently, Devaux et al.~\cite{Devaux2019} developed an “acoustic switch” device based on deforming a water-air interface, through the radiation pressure effect.  The principle is then used to achieve an efficient acoustic transmission in a specified direction of propagation but not in the opposite, hence resulting in a highly nonreciprocal transmission. One advantage of the device is that, it does not rely on nonlinear conversion of mode, hence  the transmitted wave in the forward configuration preserves most of its features, with high energy transmission. However, it suffers limitations resulting in a slow response time and the requirement of a specific orientation with gravity.} 

In this article, we address \revision{the design of an} AD composed of a stepped chain of spheres obeying Hertzian's contact law; the first chain of a given sphere size and young's modulus acts as the nonlinear medium to demodulate signals, while the second chain provides the low-pass filter made of soft spheres and equal sizes. \revision{Through the analysis of the propagation of a modulated carrier wave as function of the relevant parameters of the stepped chain, we investigate the features of the transmitted demodulated wave in relation with the excitation wave packet.}

Indeed, alignments of spheres with size gradients~\cite{Melo2006} and size steps~\cite{Job2007} proved to be an efficient shock mitigation apparatus \revision{in one direction of propagation and therefore nonreciprocal systems}. Here, we examine loaded chains, stepped in elasticity, in an attempt to break the reciprocity between nonlinear (hard) and linear (soft) media, to take advantage of low-pass filtering capabilities, \revision{and to enhance nonlinearities} to improve wave demodulation. We show that the features of the demodulated signal are strongly determined by three main effects: nonlinearity, wave dispersion, and energy dissipation. \revision{Instead of using a continuous wave excitation}, we test propagation using a wave packet with a defined envelope shape and a unique, but variable, carried frequency, \revision{aiming at exploring the whole regime of excitation frequency in the Brillouin's zone of the lattice}. We show that both high-wave dispersion and mechanical nonlinearity lead to the development of a compression front of low-frequency content that propagates at a speed corresponding to an infinite wavelength, according to the dispersion relation of the phononic lattice. The amplitude of the speed of the transmitted wave is proportional to the modulation function. However, with high mechanical nonlinearity and low wave dispersion, the bead speed of the transmitted wave is proportional to the derivative in time of the modulated function. It is shown that better information transmission is obtained in this latter regime. Thus, enhancing nonlinearities, \revision{and defining an operating frequency range near half of the cut-off frequency}, are crucial for improving performance in AD.

\begin{figure}[b]
\centering
\includegraphics[width=\textwidth]{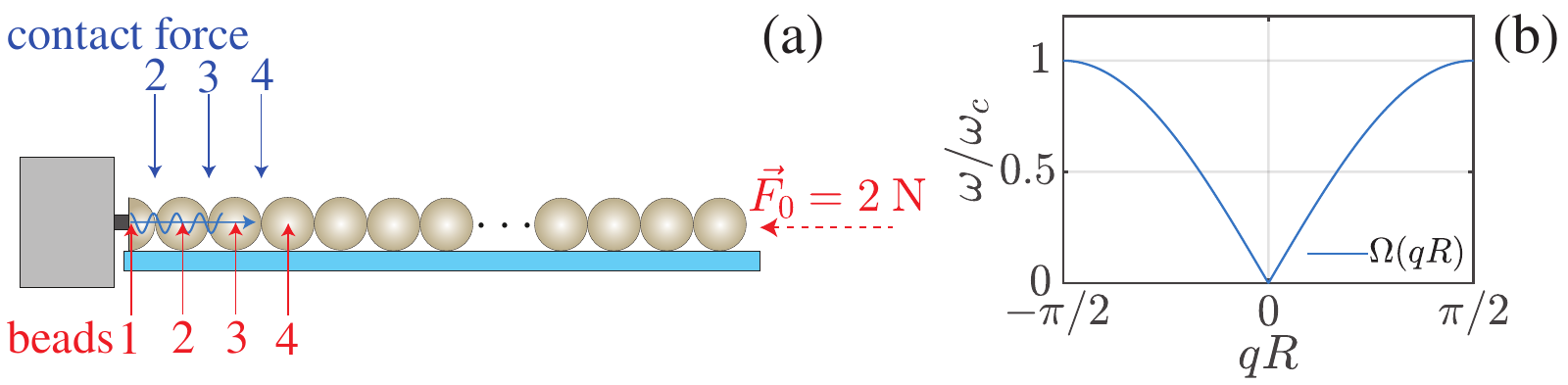}
\caption{(a) Schematic of the numerical setup for the study of wave propagation in a compressed chain excited by a mechanical actuator at one end. Blue arrows indicate the contact force between beads and red arrows the number of the beads and $\vec{F_0}$ is the compression force. \revision{(b) Dispersion relation curve $\omega(q)$, see Eq.~\ref{eq:dispersion_relation_mono}, of the alignment shown in (a), whose inflection depicts dispersive effects close to the cutoff frequency $\omega_c$ delimiting an upper forbidden band.}}
\label{fig:01_setup}
\end{figure}


\section{Numerical model}

For the efficient exploration of the main parameters controlling the nonlinear effect of the chain, we develop a numeric modeling which includes realistic experimental dimensions and beads material properties. On purpose, we consider two different materials in the present study. The parameters selected to illustrate the results correspond, first, to those of Bakelite spheres used in recent experiments~\cite{Vasconcellos2022}. This material is chosen because it was shown it is easy to machine making it possible the physical construction of a variety of composed beads~\cite{Vasconcellos2022}, compared to hard steel beads used in earlier experiments~\cite{Job2005}. Bakelite has low mass density $\rho=1775$~kg/m$^{3}$ and is nearly non-compressible with a Poisson ratio of $\nu=0.44$. In addition, although Bakelite has relatively low Young's modulus, $E=3.9$~GPa and therefore it may experience large strain, it exhibits low plastic deformation. In a previous study~\cite{Vasconcellos2022}, we measured a slightly larger value, $E=4.5$~GPa, owing to a stiffening effect due to the thwarted rotations of the particles by the friction~\cite{Job2005}. For sake of simplicity, we rule out friction in the present study so as to retain nominal estimations of Young's modulus. In addition, dissipation in Bakelite is very weak; in practice, a viscoelastic behavior with a relaxation time $\tau=4.2$~$\mu$s accurately depicts its damping features from $100$~Hz to $2500$~Hz frequency range~\cite{Vasconcellos2022}. In the present study, we also consider particles made of a material softer by more than a decade. As explained in the following, the reason is twofold: (i) its linear elastic response extends to broader ranges of driving amplitudes, and (ii) it filters lower excitations frequencies.

\begin{figure}[b]
\centering
\includegraphics[width=\textwidth]{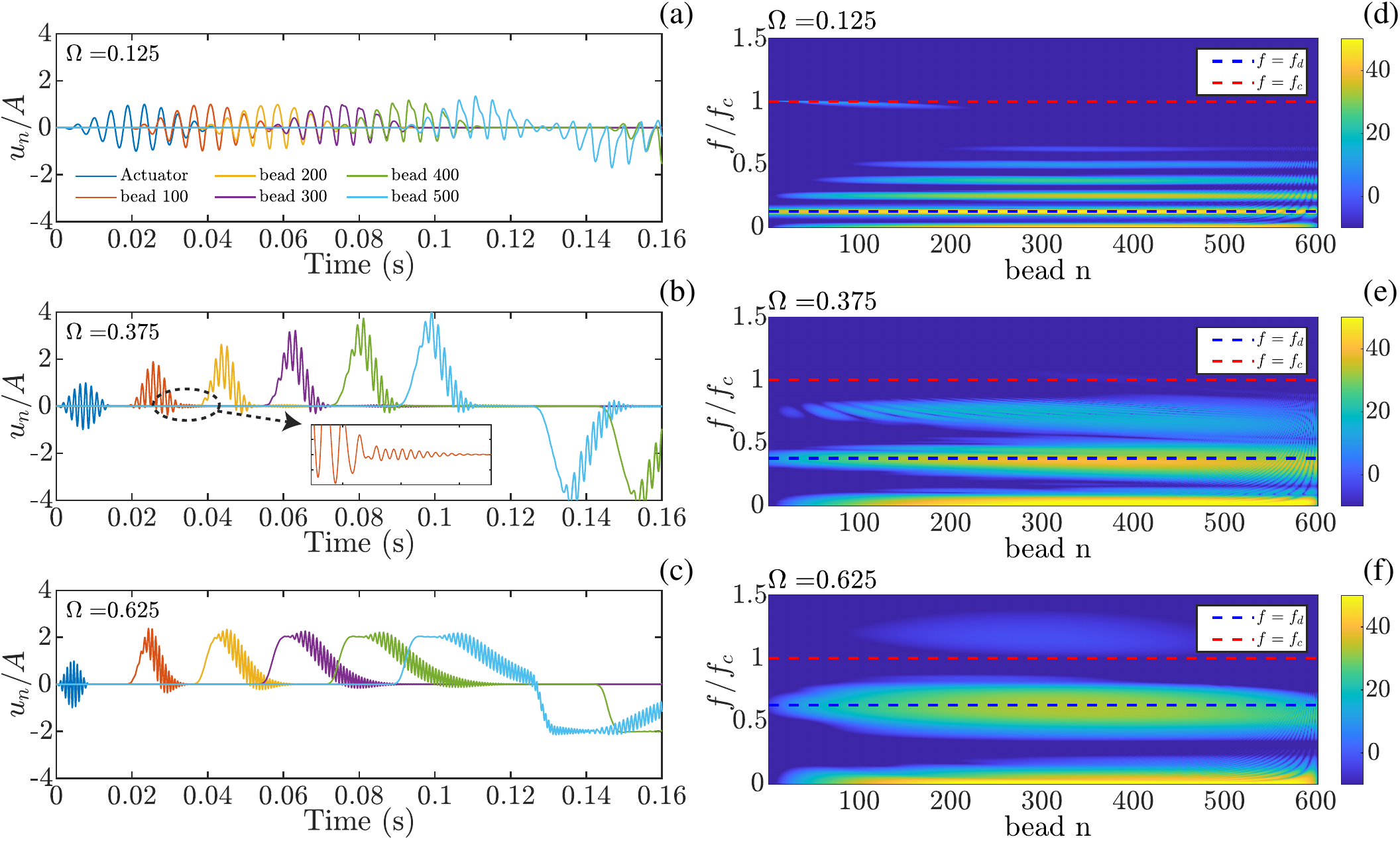}
\caption{\revision{Nonlinear waves as observed at distinct locations of the chain, at beads 100, 200, 300, 400 and 500, versus time or frequency, for $A=1$~$\mu$m. (a,d) $f_d=226$~Hz i.e. $\Omega=0.125$: harmonics progressively develop along the chain. (b,e) $f_d=677$~Hz i.e. $\Omega=0.375$: dispersion is visible through the second harmonic, which runs delayed with respect to the fundamental mode; inset depicts a magnified view of the second harmonic. (c,f) $f_d=1128$~Hz i.e. $\Omega=0.625$: a stiff front develops during the wave propagation. In all cases, wave reflection can be observed at beads 400 and 500. The spectral contents (the modulus of the single-sided amplitude spectrum $|\tilde{u}_n(\omega)|/A$ in dB) reveal the LF mode and harmonics of the carrier frequency: (d) 3rd harmonic is clearly visible, (e) 3rd harmonic is filtered since it exceed the cut-off frequency. (f) 2nd and 3rd harmonics are suppressed since their respective frequencies are beyond the cut-off frequency.}}
\label{fig:02_mono_vs_time_and_freq}
\end{figure}

In order to rule out the effect of any other parameters than elasticity, and without loss of generality, we consider here a softer but fictitious material having the same features as Bakelite except the elastic modulus, $E^\prime=E/20=195$~MPa. All our spheres have the same radius, $R=15$~mm, which leads to a particles mass $m_b=25.1$~g. Here, the alignments are made of combinations of variable number and type of beads, the first configuration probed being the monodisperse case, see Fig.~\ref{fig:01_setup}a.
The beads are aligned on a horizontal track and can be slightly compressed by an external force, $F$, along the chain. A nonlinear compression wave is initiated at one end of the chain by means of an actuator that is able to impose the displacement of the first bead of the chain. It is important to notice that this assumption suppose an actuator whose response is not affected by the load due to the chain. In addition, the boundary condition at this end for any reflected wave resulting from a reflection on the opposite end would explicitly depends on actuator response and will require further analysis, given later in the text. 
The traveling wave generated can be monitored by measuring the load with a transducer inserted inside a bead cut in two parts,~\cite{Vasconcellos2022}. The total mass of the active bead can be in practice tuned to match the mass of a regular bead and the embedded sensor thus allows non-intrusive measurements of the force across the chain. In the simulation it is simply implemented by calculating the force at the contact of two successive spheres. 

Our numerical simulation based on a Runge-Kutta algorithm explores the main features of nonlinear waves in an alignment of $N$ spheres, by solving the nonlinear system of $N$ equations, 
\begin{equation}
m\ddot{u}_{n}=\kappa\left( \delta_{0} + u_{n-1} - u_{n}\right)^{3/2} -\kappa\left(\delta_{0} + u_{n} - u_{n+1}\right)^{3/2}
\label{eq:pde_mono}
\end{equation}
for $n=(1\ldots N)$. The prefactor $\kappa=(4/3)E_{\ast}R_{\ast}^{1/2}$ depends on the reduced elastic modulus $E_{\ast}=E/2(1-\nu^2)$ and on the reduced radius of curvature $R_{\ast}=R/2$. The instantaneous displacement of particle $n$ is $u_n(t)$, and $\delta_{0}=\left( F_{0}/\kappa_{n} \right)^{2/3}$ is the static overlap resulting from the application of a static and uniform compression force $F_{0}$ along the alignment of spheres.

Due to the fact the chain is under load a dispersion relation exist in a linear regime that is found by developing all beads displacements around the corresponding static compression, this is,
\begin{eqnarray}
\Omega(q) &=& \omega/\omega_c = \sin(qR),\label{eq:dispersion_relation_mono}\\
q(\Omega) &=& \sin^{-1}(\Omega)/R = 2\pi/\lambda-j/l_{att},\nonumber\\
v_{LF} &=& \mbox{lim}_{\omega\to0}{\Re\left(\partial\omega/\partial q\right)} = \Re\left(R\omega_c\right),\nonumber\\
v_{HF} &=& \Re\left(\partial\omega/\partial q\right) = \Re\left(v_{LF}\sqrt{1-\Omega^2}\right),\nonumber
\end{eqnarray}
where $\omega=2\pi f$ is the angular frequency, $\omega_c=2\sqrt{k/m}$ is the angular cutoff frequency and $\Omega$ is the frequency normalized to the cutoff, with $k=\partial F/\partial\delta=(3/2)\kappa^{2/3}F_{0}^{1/3}$ the linear contact stiffness between particles. \revision{The dispersion relation curve $\omega(q)$ of the alignment depicted in Fig.~\ref{fig:01_setup}a is shown in Fig.~\ref{fig:01_setup}b.} The real (resp. imaginary) part of the frequency dependent wavenumber $q$ given by Eq.~\ref{eq:dispersion_relation_mono} is inversely proportional to the wavelength $\lambda$ (resp. to an attenuation length $l_{att}$). Wavelength relies on wave speed, whereas attenuation relies on dissipation (e.g. $k\in\mathbb{C}$) and/or evanescence (e.g. $\Omega>1$). In the forbidden band ($\Omega>1$), the propagation ceases (the group velocity is zero, see below) and neighbor particles oscillate in opposite phase, $\Re(q)=\pi/2R$. This gives birth to evanescent waves localized close to the actuator~\cite{Job2009,Theocharis2010}, within the localization length $l_{att}=-1/\Im(q)=R/\cosh^{-1}(\Omega)$, see Eq.~\ref{eq:dispersion_relation_mono}. In the propagating band ($0<\Omega<1$), low frequency (LF; long wavelength) signals are non-dispersive, the group velocity being $v_{LF}=\mathrm{const}$ when $\Omega\rightarrow0$ see Eq.~\ref{eq:dispersion_relation_mono}. In turn, high frequency (HF) waves are dispersive, the group velocity $v_{HF}(\omega)$ ranging from $v_{LF}$ ($\Omega\rightarrow0$) to zero ($\Omega\geq1$) see Eq.~\ref{eq:dispersion_relation_mono}. To provide external excitation at one end of the chain, a sphere of equal properties and of coordinate $u_{0}(t)$ is made to oscillate through a mechanical actuator. In all simulations, the alignments of spheres are subjected to a static compression of $F_{0}=2$~N leading to a cutoff frequency in the first Brillouin zone of $f_c=\omega_c/2\pi=1806$~Hz with $\delta_0=$3.7~$\mu$m for Bakelite particles, and $f_c^\prime=665$~Hz with $\delta_0^\prime=$27.3~$\mu$m for the softer beads.

\begin{figure}[b]
\centering
\includegraphics[width=\textwidth]{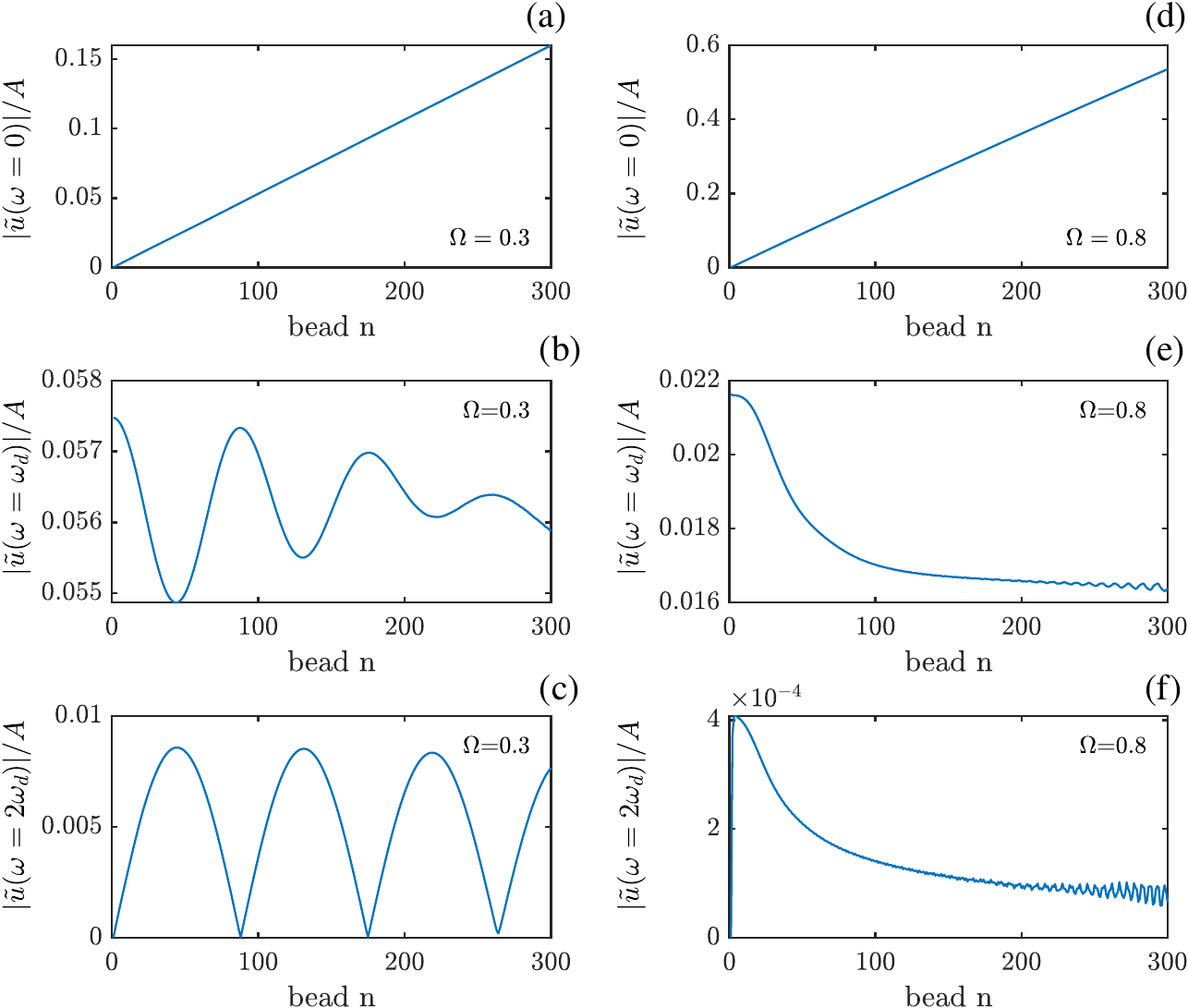}
\caption{\revision{Modulus of the single-sided amplitude spectrum $\tilde{u}_n(\omega)$ as a function of the propagation distance along the alignment of beads, when (a,b,c) the fundamental component at $\omega=\omega_d$ and the second harmonic $\omega=2\omega_d$ both lie in the propagative band ($\Omega<0.5$), and (d,e,f) when the fundamental component lies in the propagative band and the second harmonic lies in the forbidden band ($\Omega>0.5$). The LF mode increases linearly with distance, the growth rate at $\Omega=0.8$ being larger than that at $\Omega=0.3$. The fundamental component and the second harmonic oscillates spatially with a period of about 90 beads when $\Omega=0.3$, while they both decay with distance when $\Omega=0.8$.}}
\label{fig:03_mono_vs_position}
\end{figure}


\section{Parametric analysis}


\subsection{Nonlinearity}

We first investigate the nonlinear response of the chain of Bakelite spheres in the whole range of accessible frequencies.
A previous work~\cite{Sanchez2013} focused in the weak nonlinear regime of quadratic nonlinearities, demonstrating analytically the nonlinear generation of harmonics. For a continuous excitation with normalized frequency in the range $1/4<\Omega<1/2$, it was shown that the second harmonic, at twice the excitation frequency, develops with an amplitude which is modulated along the chain.
In contrast, the amplitude of the mode at vanishing frequency growths linearly with the position of the beads. For $\Omega>1/2$ large dispersion effects and the total suppression of second harmonic propagation are expected due to the cut-off frequency induced by the lattice. In the following we investigate the presence of these modes as a function of the driving frequency and excitation amplitude for the full range of frequencies and excitation amplitude. Thus, in the following the effect of the excitation frequencies in the range of $1/8<\Omega<1$ is explored for a particular excitation mode $u_0(t)$ of amplitude $A$ such that $\epsilon=A/2\delta_0\ll1$~\cite{Sanchez2013}. The form of excitation is chosen to be, 
\begin{equation}
u_{0}(t)=A\sin(\omega_d t)\times h(t)
\label{eq:imposed_motion}
\end{equation}
where $\omega_d=2\pi f_d$ is the angular driving frequency, \revision{i.e. the carrier frequency which will be denoted as the high frequency (HF) oscillation}, with $f_d=1/T_d$ its frequency and $T_d$ its period. The function $h(t)$ is a time window defined as $h(0\leq t\leq n_c T_d)=a_{0}+a_{1}\cos(\omega_d t/n_c)$ and zero otherwise, with $a_{0}=1/2$, $a_{1}=-1/2$. Thus, when changing the excitation frequency $f_d$, the envelope is adjusted such that the number of oscillations is kept constant equal to $n_c=10$ cycles. The amplitude modulation $h(t)$ thus carries the information to be transmitted in one direction and blocked in the reverse direction. Without loss of generality, we restrict our analysis to a signal with a well defined \revision{low frequency (LF)} content, peaked at $f_{LF}=f_{HF}/10$. In practice, the modulation $h(t)$ should be replaced by more complex LF information.
\revision{In the following, the LF modulation will be extracted for further analysis by using a phase preserving Butterworth low-pass digital filter of order 4 with a cutoff frequency at twice the modulation frequency, $f_\mathrm{filter}=2f_{LF}=f_{HF}/5$. Alternatively, one can also consider the LF component, $\omega=0$, of the single-sided amplitude spectrum of the signals $\tilde{u}_n(\omega)$ of the time-signal $u_n(t)$.}


\subsection{Dispersion}

We first explore the system response at low frequency, $f_d=226$~Hz, i.e., $\Omega=0.125<1/4$ at amplitude $A=1$~$\mu$m. $\epsilon=0.134$.
In line with~\cite{Fang2020}, our waveguide is designed to be long enough such that the waves reflected at the extremites do not interfere with the incident and the transmitted waves.
We thus arbitrarily consider a chain made of 602 beads; the first particle is the actuator with imposed motion and the last particle is motionless (it sustains the static compression but reflects incident waves). Fig.~\ref{fig:02_mono_vs_time_and_freq}a indicates the wave signal as seen at distinct locations in the chain. A clear development of harmonics is seen as wave packet propagate along the chain. Fig.~\ref{fig:02_mono_vs_time_and_freq}d depicts the corresponding frequency content as a function of the bead position. Several nonlinear modes, $2f$, $3f$ and $4f$, with amplitude depending on the position on the chain are clearly visible. 

Increasing the excitation frequency in the range $1/4<\Omega=0.375<1/2$ reveals dispersion effects. These dispersive effects are very prominent in the second harmonic as its frequency becomes closer to the limit of the Brillouin zone. The amplitude of second harmonic is relatively small, see Fig.~\ref{fig:02_mono_vs_time_and_freq}b, and propagates at a speed that is lower than that of the fundamental mode. Thus, the second harmonic is found completely separated from the excitation as seen at the tail of the main excitation signal, at sufficiently long distances from the source, Fig.~\ref{fig:02_mono_vs_time_and_freq}b. Accordingly, the frequency content, Fig.~\ref{fig:02_mono_vs_time_and_freq}e, indicates a second harmonic of significantly reduced amplitude.
When $\Omega$ is increased further, for $1/2<\Omega=0.625$, ($f_d=1128$~Hz) a compression front develops progressively as the excitation propagates along the chain, see Fig.~\ref{fig:02_mono_vs_time_and_freq}c. As expected the second harmonic no longer propagates because its frequency exceeds the maximum frequency allowed. Spectral content of the propagative wave, see Fig.~\ref{fig:02_mono_vs_time_and_freq}f, as function of distance from the source indicates that the amplitude of LF mode develops progressively with propagation along the chain. 

In order to study in more detail the amplitude of these modes, we compute the average amplitude detected at every contact. The spectral amplitude of the LF mode increases linearly with bead position, see Figs.~\ref{fig:03_mono_vs_position}a and~\ref{fig:03_mono_vs_position}d, whereas those of the fundamental component and the second harmonic are periodic in space, see Figs.~\ref{fig:03_mono_vs_position}b and~\ref{fig:03_mono_vs_position}c, when both lie in the propagative band. In turn, the fundamental component and the second harmonic both decay with distance, by no more $20$\% for the former and by $80$\% for the latter, when the second harmonic falls in the forbidden band. These results are consistent with earlier theoretical predictions by Sanchez et al.~\cite{Sanchez2013} obtained in the case of a continuous excitation. \revision{The slight decay of the fundamental amplitude observed at large distance of propagation when $\Omega>0.5$ is a combined effect of both the nonlinearity and the dispersion, associated with the spatially compact, thus finite energy, wave packet excitation in our case, whereas the large decay of the second harmonic stems from its evanescent nature.}

\begin{figure}[b]
\centering
\includegraphics[width=\textwidth]{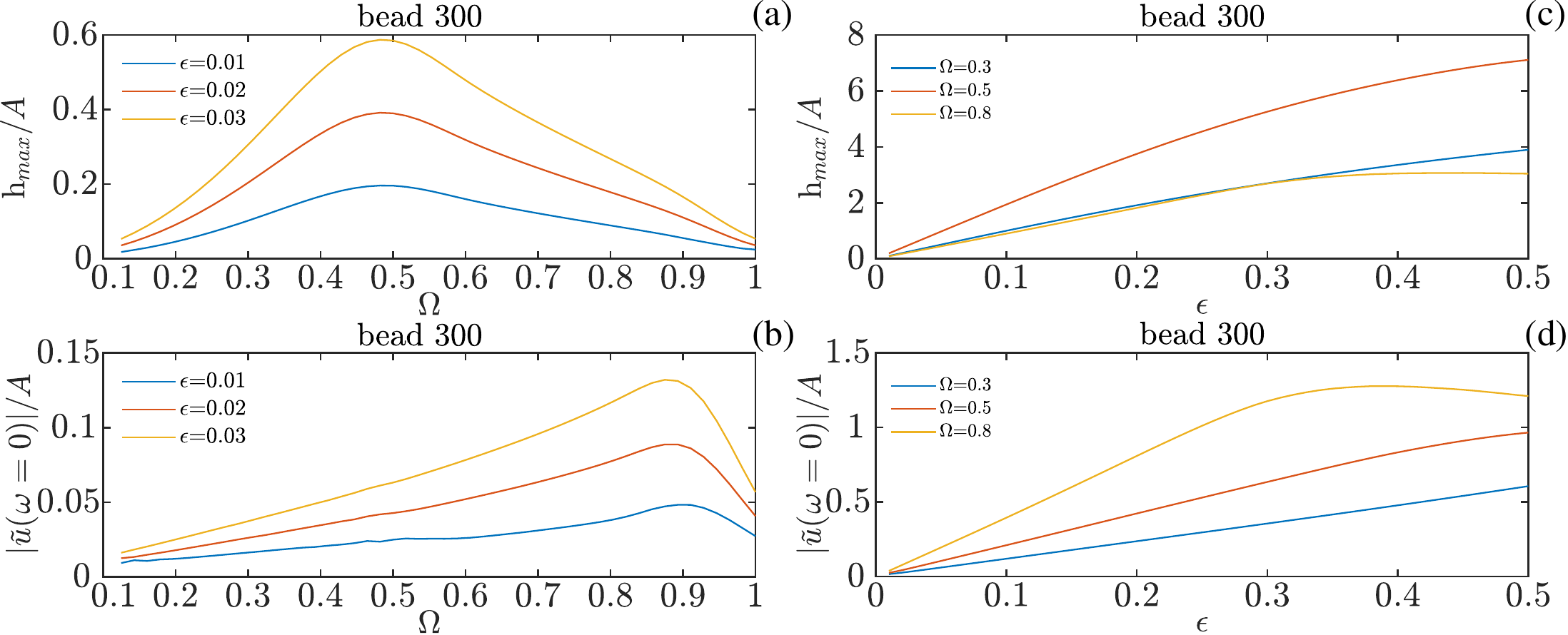}
\caption{\revision{(a) Stationary amplitude of the low frequency compression front, $h_{max}$, and (b) modulus of the single-sided amplitude spectrum of the LF mode, $|\tilde{u}(\omega=0)|$, as function of the reduced excitation frequency, $\Omega$, for several excitation amplitude, $\epsilon$. (c) $h_{max}$ and (d) $|\tilde{u}(\omega=0)|$ versus $\epsilon$ for various $\Omega$.}}
\label{fig:04_mono_vs_omega_and_epsilon}
\end{figure}

In the following the features of the compression front generated in the high dispersion regime is investigated as function of $\Omega$ and the excitation amplitude $\epsilon$. The amplitude of this front, $h_{max}$ and its connection with the frequency content of the LF mode is first explored. Figure~\ref{fig:04_mono_vs_omega_and_epsilon}a demonstrates that $h_{max}$ has a maximum near $\Omega\approx 0.5$, which increases with the excitation, $\epsilon$. This maximum is attributed to the the fact that for small and increasing $\Omega$, the excitation envelope grows faster, but at sufficiently high values of $\Omega$ the dispersive effect dominates producing a spacial separation of the front and its tail, see Fig.~\ref{fig:02_mono_vs_time_and_freq}c. By contrast, the amplitude of the LF mode has a maximum that is close to $\Omega\approx 0.8$, and the value of this maximum increases with $\epsilon$, see Fig.~\ref{fig:04_mono_vs_omega_and_epsilon}b. The location of this maximum indicates that the amplitude of LF mode continue to growth through the development of the flat part of the signal. The decrease observed as $\Omega$ approaches 1 results from the fact that the excitation signal, although it is centered at a frequency smaller than the cut-off one, has a bandwidth where the highest frequencies overcome this cut-off.

The detailed variation of these quantities with excitation amplitude $\epsilon$ is presented in Figs.~\ref{fig:04_mono_vs_omega_and_epsilon}c and~\ref{fig:04_mono_vs_omega_and_epsilon}d. The rapid increase of the LF mode with the excitation amplitude is corroborated with early predictions in the weakly nonlinear regime~\cite{Sanchez2013}. However, for high enough $\Omega$ dispersion is more prominent leading to reduced effect of nonlinearities and to a LF mode that grows at lower rates.


\subsection{Wavefront separation and saturation}

\begin{figure}[b]
\centering
\includegraphics[width=0.75\textwidth]{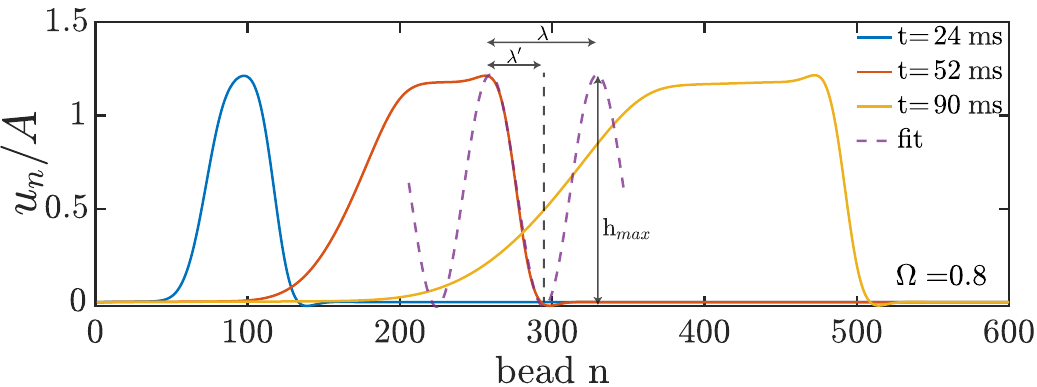}
\caption{Beads displacement at successive propagation instants, for $\Omega=0.8$. The induced compression remains unchanged whereas the relaxation tail becomes smoother with propagation. The front size is characterized through a typical length, $\lambda'$ from Fourier's series with argument $(2\pi x/\lambda)$, where $\lambda'=\lambda/2$.}
\label{fig:05_mono_wavefront_measure}
\end{figure}

As discussed previously, one of the main features of the nonlinear regime, in the highly dispersive limit (large $\Omega$), is the development of a front characterized by a sharp bead displacement of maximum amplitude, $h_{max}$. In order to understand what limits $h_{max}$, the interplay between the nonlinear and the dispersion effects is investigated. It is first observed that in addition to the compression front a slow decompression takes place. We first characterize the width in space of such low frequency signal through the filtering of the high frequency signal whose frequency content is dominated by the excitation frequency. The low pass filtered wave reveals that the form of the compression front remains unchanged with propagation, whereas the decompressing tail slowly grows in width, see Fig.~\ref{fig:05_mono_wavefront_measure}. The compression front can be approximated by a segment of a sine function of width $\lambda'$, see Fig.~\ref{fig:05_mono_wavefront_measure}.

\begin{figure}[b]
\centering
\includegraphics[width=0.75\textwidth]{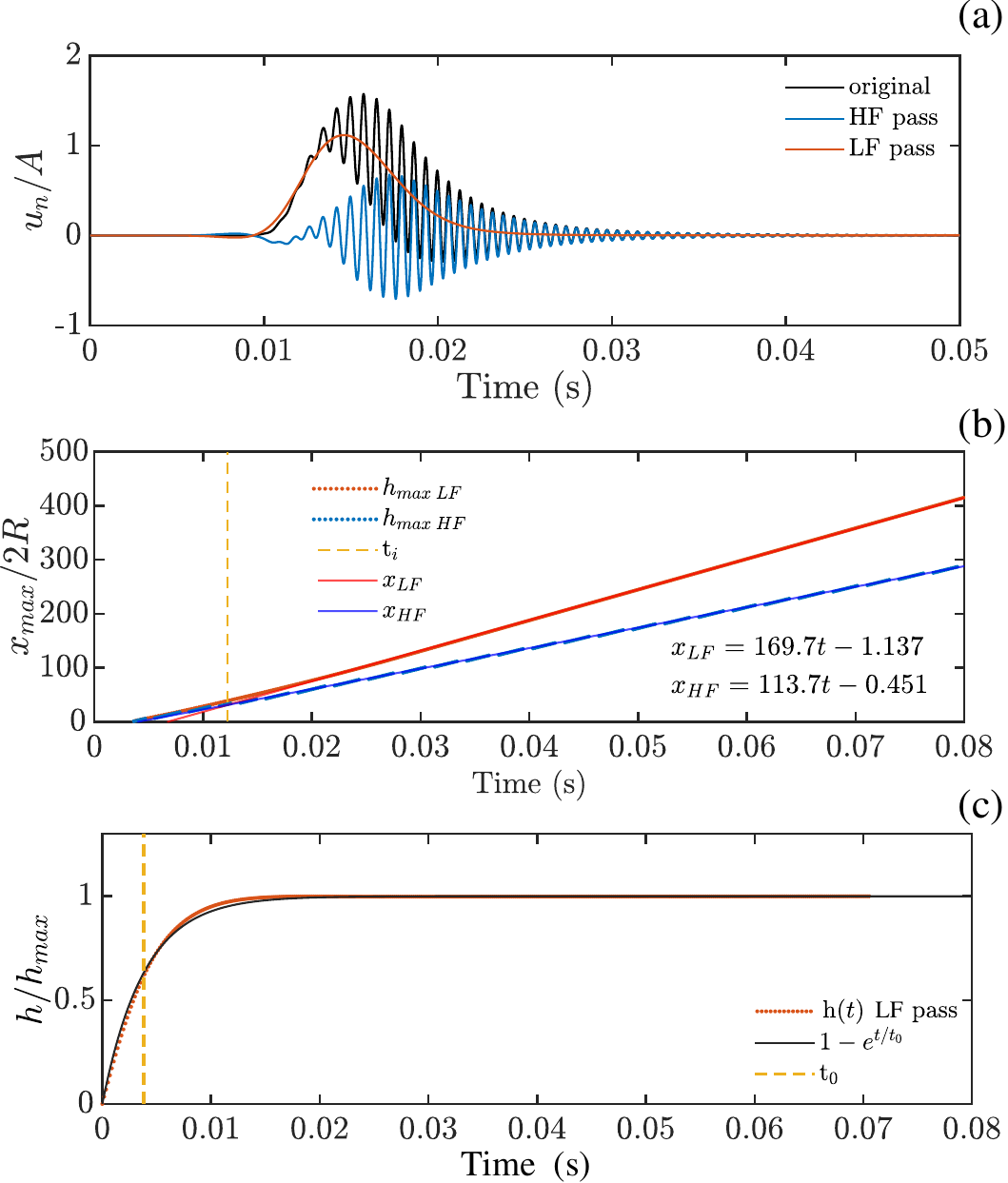}
\caption{a) The full wave observed at bead 50 for $\Omega=0.8$. LF and HF wave packet. b) $x_{max}$ is the distance of $h_{max}$ from the actuator. The positions of both maxima, of the compression front $x_{LF}$ and the wave packet $x_{HF}$ as function of time. c) The maximum of the compression front as function of time. An exponential fit with characteristic time $t_{0}$ captures the front dynamics.}
\label{fig:06_mono_HF_LF_waves}
\end{figure}

Dispersive effects are present at sufficiently high $\Omega$ and are responsible of compression front separation. In order to account for these effects, the propagation speed of both, the front and the wave packet centered in the excitation frequency are investigated. Through the application of a low pass filtering and through the subtraction of the low frequency profile to the whole signal, the desired modes can be obtained, see Fig.~\ref{fig:06_mono_HF_LF_waves}a. A significant delay of the maximum of the developing front and the maximum of the wave packet is observed. Following such delay in time, see Fig.~\ref{fig:06_mono_HF_LF_waves}b, indicates that compression does not start to develop instantaneously and therefore is at early stages running behind the wave packet. However, front speed is higher than that of wave packet which leads to a crossing point, $t_{int}$ that is indicated in Fig.~\ref{fig:06_mono_HF_LF_waves}b. This speed difference is consistent with the relation dispersion of the chain which predicts a group speed that decreases with $f$, Eq.~\ref{eq:dispersion_relation_mono}.

This analysis leads to address the question of what determines the maximum growth of the compression front. A simple argument is that the speed difference multiplied by the elapsed time required to produce a distance difference equals to the width of the front. This leads to an expression for the widh of the front, as follows,
\begin{equation}
\lambda'=C(v_{LF} - v_{HF})(t_{0}-t_{i})
\label{eq:wavefront_length}
\end{equation}
where $v_{LF}$ and $v_{HF}$ are the LF front speed and the HF group speed of HF wave packet, $t_{0}$ is the typical growth time of the front, $t_{i}$ is the instant at which the LF and th HF components crossover each other, see Fig.~\ref{fig:06_mono_HF_LF_waves}b, and $C$ is a proportionality constant.

\begin{figure}[b]
\centering
\includegraphics[width=0.75\textwidth]{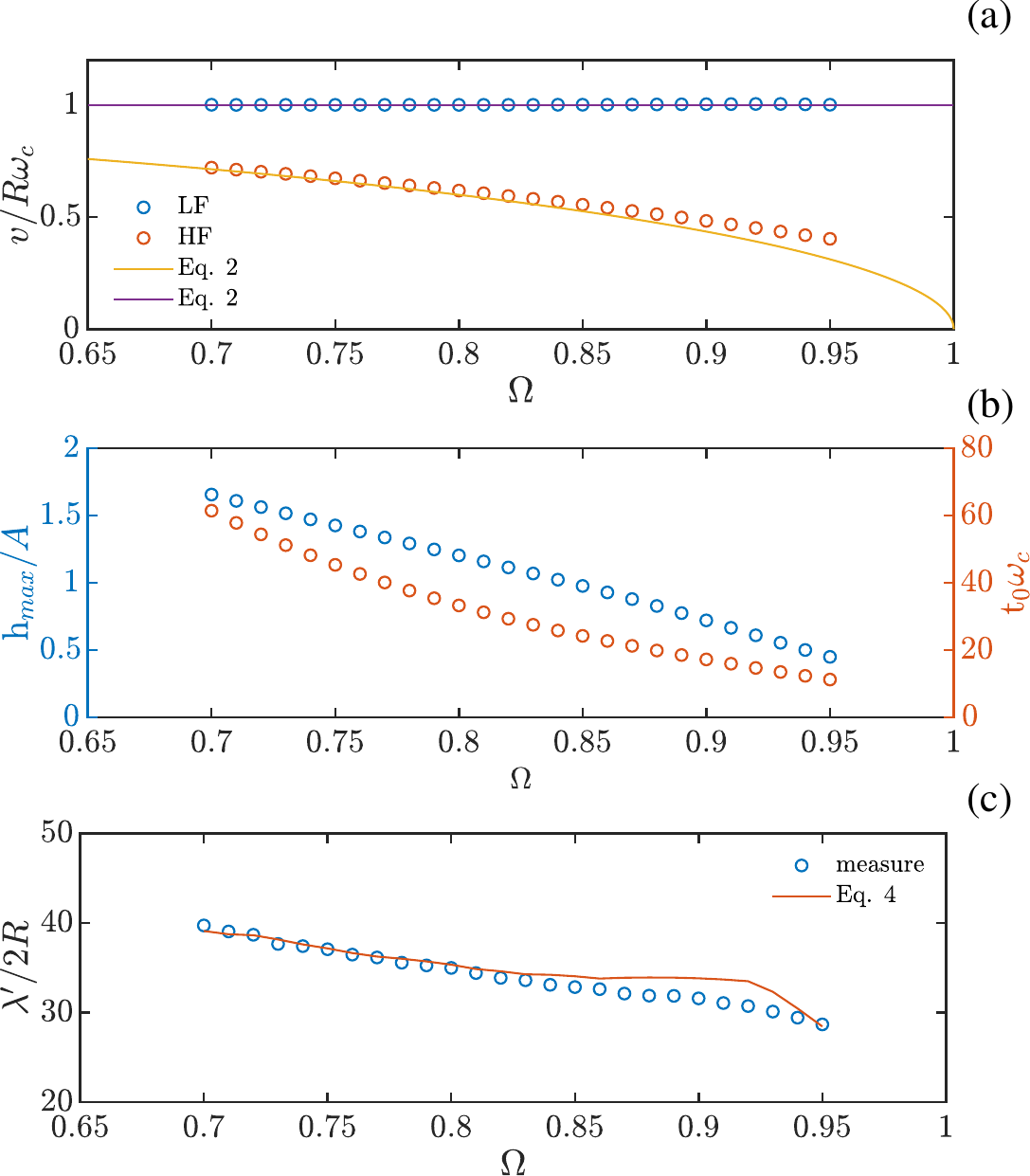}
\caption{ a) LF front speed $v_{LF}$ and HF group speed $v_{HF}$ of the wave packet as function of $\Omega$ (solid lines are predictions from Eq.~\ref{eq:dispersion_relation_mono}). b) The maximum amplitude of the compression front $h_{max}$ and characteristic time $t_{0}$ vs $\Omega$. c) Size of the compression front, $\lambda'$ vs $\Omega$. Solid line is the geometrical relationship of the front parameters, as proposed by Eq.~\ref{eq:wavefront_length}. $C=-1.35$, $t_{i}>t_{0}$.}
\label{fig:07_mono_wavefront_model}
\end{figure}

In order to determine $t_{0}$ and validate the above expression (Eq.~\ref{eq:wavefront_length}), through the filtering procedure the maximum amplitude of the compression time is followed in time, see Fig.~\ref{fig:06_mono_HF_LF_waves}c. A saturating exponential nearly describes the time-evolution of the compression front, which can be approximated by $1-e^{-t/t_{0}}$, where $t_{0}$ is interpreted as the typical growth time. Systematic measurements of the waves speed, front width and characteristic time as function of $\Omega$ leads to data on Fig.~\ref{fig:07_mono_wavefront_model}. Waves speed extracted by filtering follow the speed predicted by the linear relation dispersion Eq.~\ref{eq:dispersion_relation_mono}. Group speed is nearly constant for the low frequency front, regardless the excitation frequency whereas, large dispersion is observed for the wave packet when $\Omega$ tends to 1, see Fig.~\ref{fig:07_mono_wavefront_model}a. Figure~\ref{fig:07_mono_wavefront_model}b accounts for the maximum particle displacement involved in the compression front, $h_{max}$ and the characteristic time necessary to reach this amplitude. 

The front width decreases with $\Omega$, in fact, this dependence is attributed to the fact that the excitation envelope decreases with $\Omega$ since the number of oscillations is maintained constant. Finally, the scaling for the characteristic front width proposed in Eq.~\ref{eq:wavefront_length} is verified in Fig.~\ref{fig:07_mono_wavefront_model}c. 


\subsection{Nonlinear demodulation} 

\begin{figure}[b]
\centering
\includegraphics[width=\textwidth]{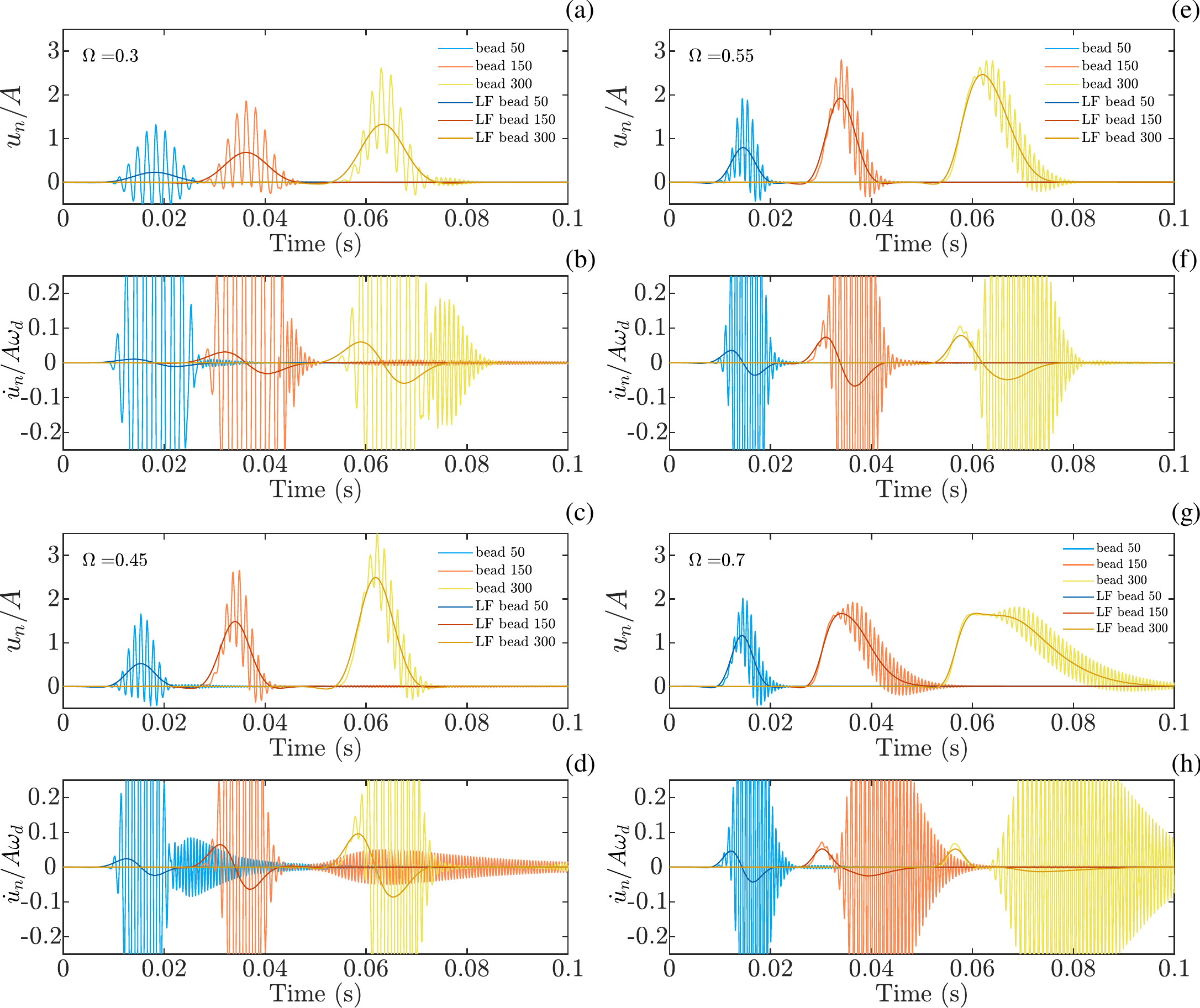}
\caption{\revision{(a,c) Wave packets of particles displacements propagating along the chain for low values of $\Omega<0.5$ at distinct locations in the chain. The low frequency contribution of the particles displacement obtained from low pass filtering are shown in thick solid lines. (b,d) Particles velocity at distinct locations along the chain, for low values of $\Omega<0.5$, superimposed on their low frequency contribution. (e,g) Wave packets of particles displacements propagating along the chain for high values of $\Omega>0.5$, superimposed on their low frequency contribution. (f,h) Particles velocity at distinct locations along the chain, for high values of $\Omega>0.5$, superimposed on their low frequency contribution.}}
\label{fig:08_mono_demodulation_LF_and_HF}
\end{figure}

The influence of the dispersion, nonlinearity and the distance traveled by the excitation on the demodulated signal is studied in the following. This mode is obtained by low pass filtering the whole signal. 
At weak wave dispersion $\Omega<0.5$ it is observed that the demodulated particle displacement is proportional to the envelope $h(t)$, see Figs.~\ref{fig:08_mono_demodulation_LF_and_HF}a and~\ref{fig:08_mono_demodulation_LF_and_HF}c, while the velocity results proportional to the temporal derivative of the envelope function $h'(t)$, see Figs.~\ref{fig:08_mono_demodulation_LF_and_HF}b and~\ref{fig:08_mono_demodulation_LF_and_HF}d.
However, wave dispersion has an accumulative effect on the wave shape which becomes significant for $\Omega>0.5$. Thus, the bead displacement and velocity signal observed at beads located far from the source deviates from $h(t)$ and $h'(t)$ respectively. This deviation becomes more significant at higher values of $\Omega$, see Fig.~\ref{fig:08_mono_demodulation_LF_and_HF}(e-h). Indeed, for $\Omega>0.5$, and far from the source the particle speed becomes instead proportional to the wave envelope $h(t)$, see Fig.~\ref{fig:08_mono_demodulation_LF_and_HF}(e-h). Nevertheless, it is important to notice that at short times or equivalently, at distances sufficiently close to the wave source, particle velocity remains proportional to $h'(t)$. These results indicate that the effect of dispersion leads to a transition of the demodulated particle velocity that is first proportional to $h'(t)$ at short times or sufficiently close to the source and then becomes proportional to $h(t)$ far from the source.


\subsection{Viscoelastic dissipation}

\begin{figure}[b]
\centering
\includegraphics[width=\textwidth]{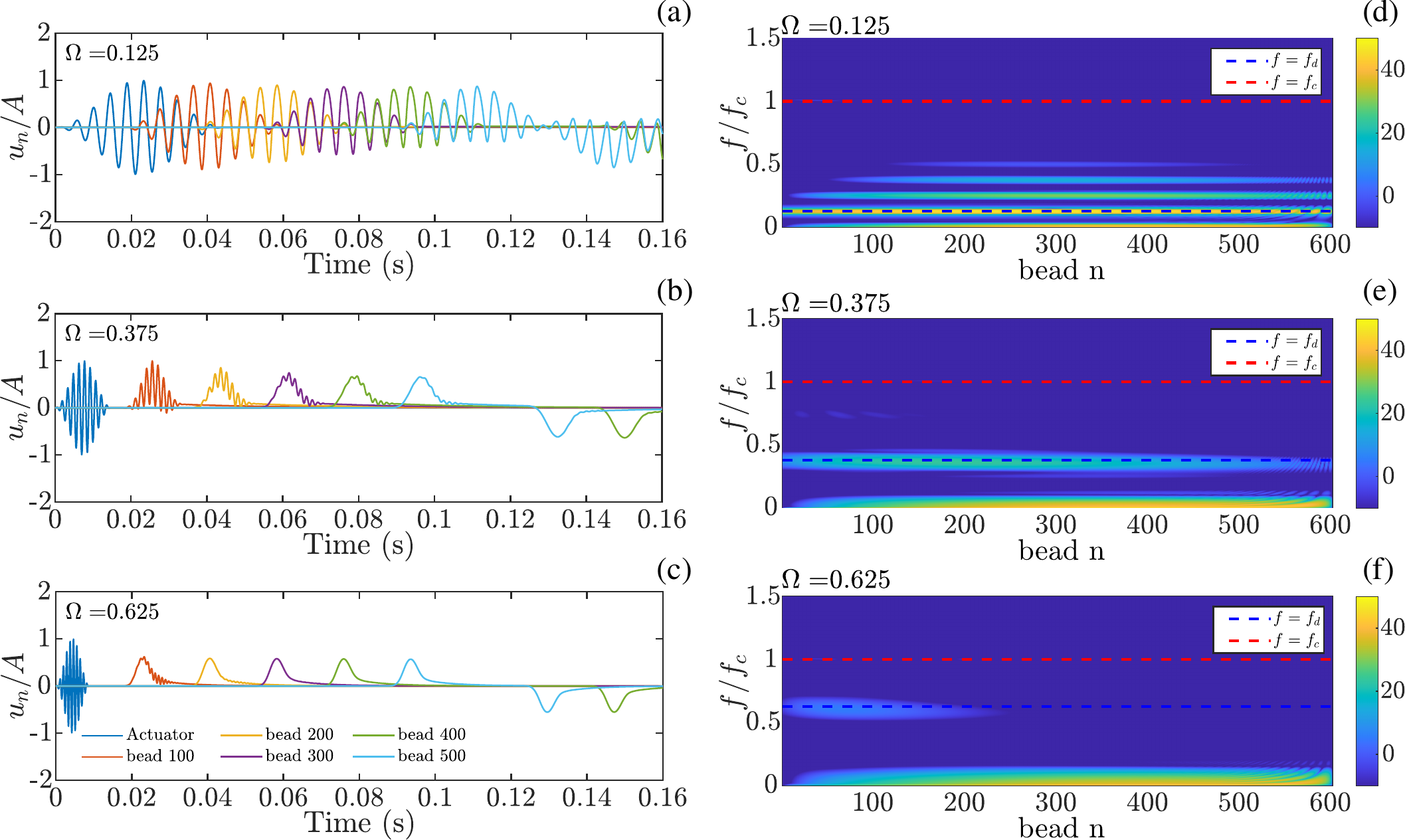}
\caption{\revision{Nonlinear waves as observed at distinct locations of the chain, at beads 100, 200, 300, 400 and 500, versus time or frequency for $A=1$~$\mu$m and accounting for dissipation ($\tau=4.2$~$\mu$s).(a,d) $f_d=226$~Hz, $\Omega=0.125$. (b,e) $f_d=677$~Hz, $\Omega=0.375$. (c,f) $f_d=1128$~Hz, $\Omega=0.625$. Panels (d,e,f) show the modulus of the single-sided amplitude spectrum $|\tilde{u}_n(\omega)|/A$ in dB.}}
\label{fig:09_mono_diss_vs_time_and_freq}
\end{figure}

The previous sections have not considered wave attenuation due to energy dissipation. Here we account for the viscous dissipation due to the material loss following the approach introduced early by Kuwabara~\cite{Kuwabara1987}. This approximation has been shown to correctly describe experimental configurations dealing with wave propagation along chains of spheres made of steel and Bakelite~\cite{Vasconcellos2022}. The dissipation due to the internal viscoelastic behavior of the bulk material expresses as, $F_D=\tau\times \partial_{t}F_H$, with $F_H=\kappa\delta^{3/2}$~\cite{Kuwabara1987} the Hertzian elastic force, $\delta$ the overlap deformation, and $\tau$ the relaxation time associated to the viscous damping of Bakelite spheres. With this, the dynamical equation writes,\\

\begin{eqnarray}
m\ddot{u}_{n} &=& \kappa\left( \delta_{0} + u_{n-1} -u_{n} \right)^{3/2} \label{eq:pde_mono_diss} \\ 
&-& \kappa \left( \delta_{0} + u_{n} - u_{n+1} \right)^{3/2}\nonumber \\
&+& (3/2)\tau\kappa \left(\delta_{0}+u_{n-1}-u_{n}\right)^{1/2} \left(\dot{u}_{n-1} - \dot{u}_{n}\right) \nonumber \\
&-& (3/2)\tau\kappa \left(\delta_{0}+u_{n}-u_{n+1}\right)^{1/2} \left(\dot{u}_{n} - \dot{u}_{n+1}\right). \nonumber
\end{eqnarray}

For realistic values of viscous parameter $\tau$ (Bakelite), the effect of dissipation on the wave packet propagation is a significant attenuation of the driving signal and its harmonics, see Fig.~\ref{fig:09_mono_diss_vs_time_and_freq}. Thus, the presence of the demodulated wave is rapidly revealed by energy dissipation. For both low excitation frequencies and dispersion, see Fig.~\ref{fig:09_mono_diss_vs_time_and_freq}a, dissipation leads to a gradual attenuation of the high frequency content. Figure~\ref{fig:09_mono_diss_vs_time_and_freq}d depicts the corresponding frequency content. As previously, harmonic and LF modes are developed with little effect of dissipation for the low frequency signals. However, as excitation frequency increases the HF content of the wave packet decreases gradually along the chain, revealing rapidly the demodulated signal, see Figs.~\ref{fig:09_mono_diss_vs_time_and_freq}b and~\ref{fig:09_mono_diss_vs_time_and_freq}c. Figures~\ref{fig:09_mono_diss_vs_time_and_freq}e and~\ref{fig:09_mono_diss_vs_time_and_freq}f show how each frequency component decrease with distance. Overall, for realistic values of viscous dissipation and frequency cut-off, dissipation does not affect significantly the demodulated wave. In the next section, we will explore how these finding can be applied to optimize the simplest nonlinear device, the acoustic diode.


\section{Implementation of an acoustic diode} 

The simplest acoustic diode can be implemented through a stepped chain made of two segments of beads of either distinct size but equal elastic properties or equal size but distinct elastic properties. In the forward configuration, one chain acts as demodulating nonlinear device while the other is a suitable low pass filter aimed to eliminate the carrier wave. While, in the backward direction, the signal impinges first the low pass filter and the carrier wave is rejected. 


\subsection{Stepped chain}

\begin{figure}[p]
\centering
\includegraphics[width=\textwidth]{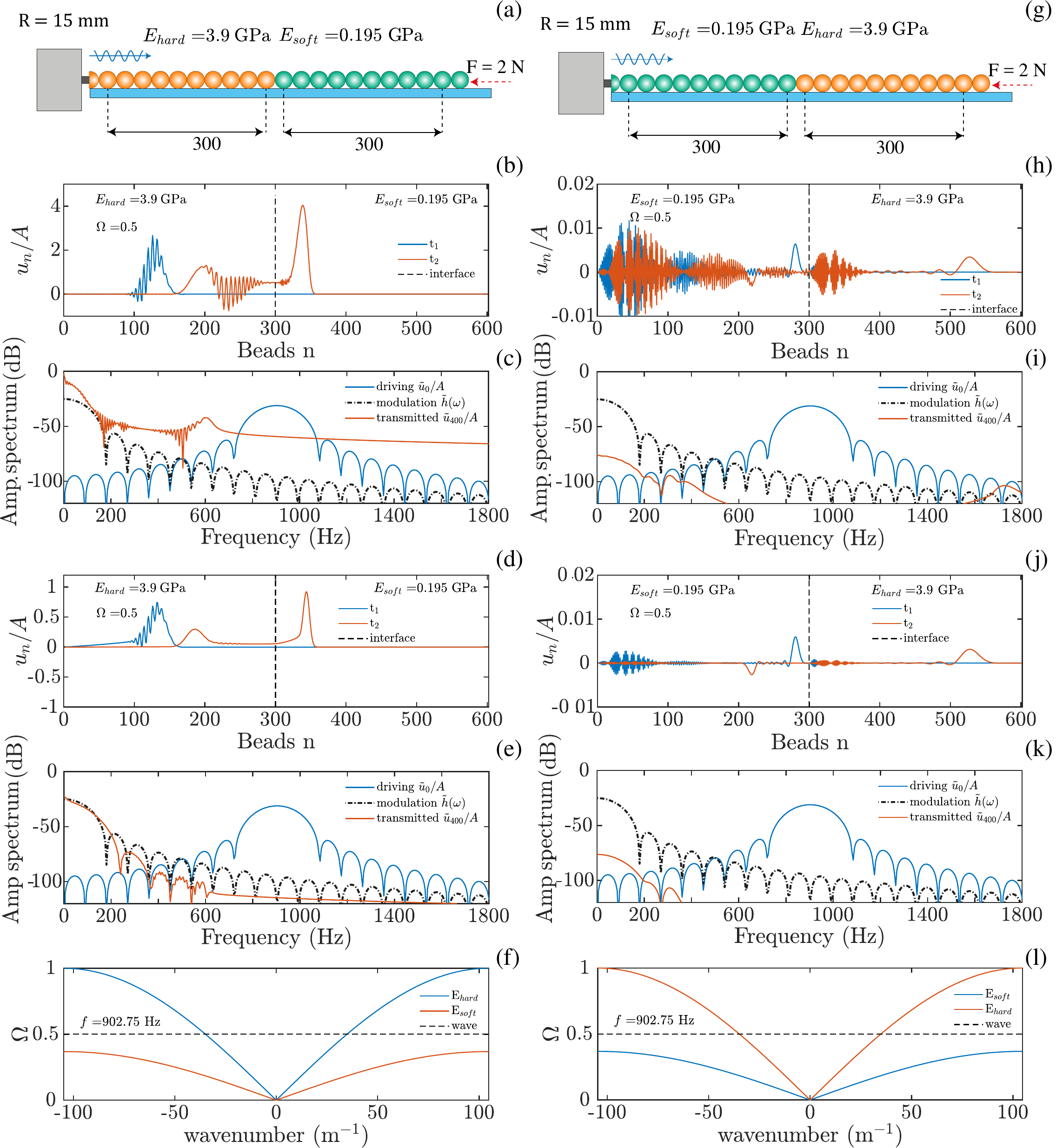}
\caption{\revision{Acoustic diode in the forward (a-f) and the backward (g-l) directions, made of two sections of distinct elasticity, at driving frequency $f_d=f_c/2=903$~Hz. Signals versus time at two distinct instants, $t_{1}=30$~ms (before transmission) and $t_{2}=80$~ms (after transmission), without (b,h) and with (d,j) viscoelastic losses. In the forward direction, the demodulated LF wave passes through the interface whereas the carrier frequency is reflected. In the backward direction, the signal driven by the actuator is filtered by the linear soft chain, the small leakage being due to a weak remaining nonlinear demodulation. (c,e,i,k) are the modulus of the single-sided amplitude spectrum $|\tilde{u}_n(\omega)|/A$ in dB of the signals $u_n(t)/A$ shown in (b,d,h,j) respectively. (f,l) are the dispersion relations in the forward and the backward directions, respectively.}}
\label{fig:10_diode_fwd_bwd}
\end{figure}

In the following we consider a stepped chain as the simplest implementation of an acoustical diode. The general equation for a chain of distinct beads writes,
\begin{eqnarray}
m_n\ddot{u}_{n}&=&\kappa_{n-1,n}\left( \delta_{n-1,n} + u_{n-1} - u_{n} \right)^{3/2} \label{eq:pde_stepped} \\ 
&-& \kappa_{n,n+1} \left( \delta_{n,n+1} + u_{n} - u_{n+1} \right)^{3/2} \nonumber
\end{eqnarray}
where $m_n=(4/3)\pi R_n^3\rho_n$ is the mass of a sphere given its radius $R_n$ and mass density $\rho_n$,
and $\kappa_{n,n+1}=(4/3)E_{n,n+1}R_{n,n+1}^{1/2}$ with $E_{n,n+1}=[(1-\nu_{n}^2)/E_{n}+(1-\nu_{n+1}^2)/E_{n+1}]^{-1}$ the reduced elastic modulus and $R_{n,n+1}=[1/R_{n}+1/R_{n+1}]^{-1}$ the reduced radius of curvature, and where $\delta_{n,n+1}=\left( F_{0}/\kappa_{n,n+1} \right)^{2/3}$ is the local static overlap resulting from the application of a static and uniform compression force $F_{0}$ along the alignment of spheres.
For the stepped chain, we consider a segment of 300 beads made of Bakelite and aligned with a segment of 300 softer particles, see Fig.~\ref{fig:10_diode_fwd_bwd}a. All contacts of a segment are thus identical except the contact at the boundary between hard and soft segments. The principle of functioning of the diode in the forward (passing) direction is depicted in  Fig.~\ref{fig:10_diode_fwd_bwd}b without dissipation and in Fig.~\ref{fig:10_diode_fwd_bwd}d with a viscoelastic dissipation involved. \revision{The frequency content of the driving excitation, the amplitude modulation, and the transmitted demodulated wave are represented in Figs.~\ref{fig:10_diode_fwd_bwd}c and~\ref{fig:10_diode_fwd_bwd}e, respectively with and without dissipation; in particular, the spectrum of the transmitted demodulated wave has to be compared to the spectrum of the amplitude modulation.} Since the chain of hard spheres allows for the propagation of a wider range of frequencies than that of the chain of softer spheres, if the carrier frequency is chosen above the cutoff frequency of the soft beads chain, the latter acts as a low pass filter, see Fig.~\ref{fig:10_diode_fwd_bwd}f. The wave propagating from left (hard) to right (soft), a self demodulated wave progressively develops in the former segment due to nonlinear effects.
Indeed, the propagation in the the hard segment is nonlinear since the amplitude of the driving displacement is the order of the static compression, $A=1~\mu\mbox{m}\sim\delta_0=3.7~\mu\mbox{m}$. In turn, the propagation in the softer segment more linear, since $A\ll\delta_0^\prime=27.3~\mu\mbox{m}$.
The envelop wave demodulated in the nonlinear segment being of low frequency, it can then propagate through the chain of softer beads. By contrast, in the backward (blocked) direction, the excitation traveling from right (soft) to left (hard) is rejected because of the filtering by the first segment, see Fig.~\ref{fig:10_diode_fwd_bwd}(h-l). Thus, the stepped chain acts as a diode allowing the transmission of the LF demodulated wave only in the forward direction and the rejection of the HF and LF signals in the backward direction.
Looking into the details of the propagation in the backward direction, a weak LF signal is transmitted in practice, akin to a current leak in a conventional diode. Despite the soft chain is less subject to demodulation compared to the more nonlinear hard chain, and despite it filters a substantial part of the incident wave, some weak demodulation occurs anyhow.
However, the isolation level of the mechanical diode under scrutiny is of the order of $60$~dB without dissipation; the amplitude of the transmitted pulse in the forward direction being about $u\simeq4A\simeq4$, see the LF transmitted pulse located between beads $300$ and $400$ in Fig.~\ref{fig:10_diode_fwd_bwd}b, and $u\simeq 0.005*A$ in the backward direction, see the pulse located between beads $500$ and $600$ in Fig.~\ref{fig:10_diode_fwd_bwd}h.
Also, for the set of probed parameters, see Fig.~\ref{fig:10_diode_fwd_bwd}, the demodulated wave is proportional to the driving signal envelope, and therefore information can be coded through the modulation of such envelope. However, if wave dispersion is high or the wave propagates long distances, the demodulated wave is no longer proportional to the envelope, which render the transmitted information more difficult to decode.


\subsection{Diode efficiency with losses}

In the numerical simulation, for the diode configuration the viscous dissipation can be generalized, as follow,\\
\begin{eqnarray}
m_n\ddot{u}_{n} &=& \kappa_{n-1,n}\left( \delta_{n-1,n} + u_{n-1} -u_{n} \right)^{3/2}\label{eq:pde_stepped_diss} \\ 
&-& \kappa_{n,n+1} \left( \delta_{n,n+1} + u_{n} -u_{n+1} \right)^{3/2} \nonumber \\
&+& (3/2)\tau_{n-1,n} \kappa_{n-1,n} \left(\delta_{n-1,n} + u_{n-1} - u_{n}\right)^{1/2} \left(\dot{u}_{n-1} - \dot{u}_{n}\right) \nonumber \\
&-& (3/2)\tau_{n,n+1} \kappa_{n,n+1} \left(\delta_{n,n+1} + u_{n} - u_{n+1}\right)^{1/2} \left(\dot{u}_{n} - \dot{u}_{n+1}\right) \nonumber
\end{eqnarray}
where $\tau_{n,n+1}$ is a relaxation time, depending on the properties of both materials in contact, which can be approximated as
\begin{equation}
\frac{\tau_{n,n+1}}{E_{n,n+1}} \simeq \left(\frac{1-\nu_{n}^2}{E_{n}}\right)\tau_{n} + \left(\frac{1-\nu_{n+1}^2}{E_{n+1}}\right)\tau_{n+1},
\end{equation}
in the weakly dissipative limit, $\omega\tau_{n}\ll1$.
Without loss of generality, we arbitrarily account for the same relaxation time in all materials, such that $\tau_{n-1,n}=\tau_{n,n+1}=\tau$.

\begin{figure}[b]
\centering
\includegraphics[width=0.75\textwidth]{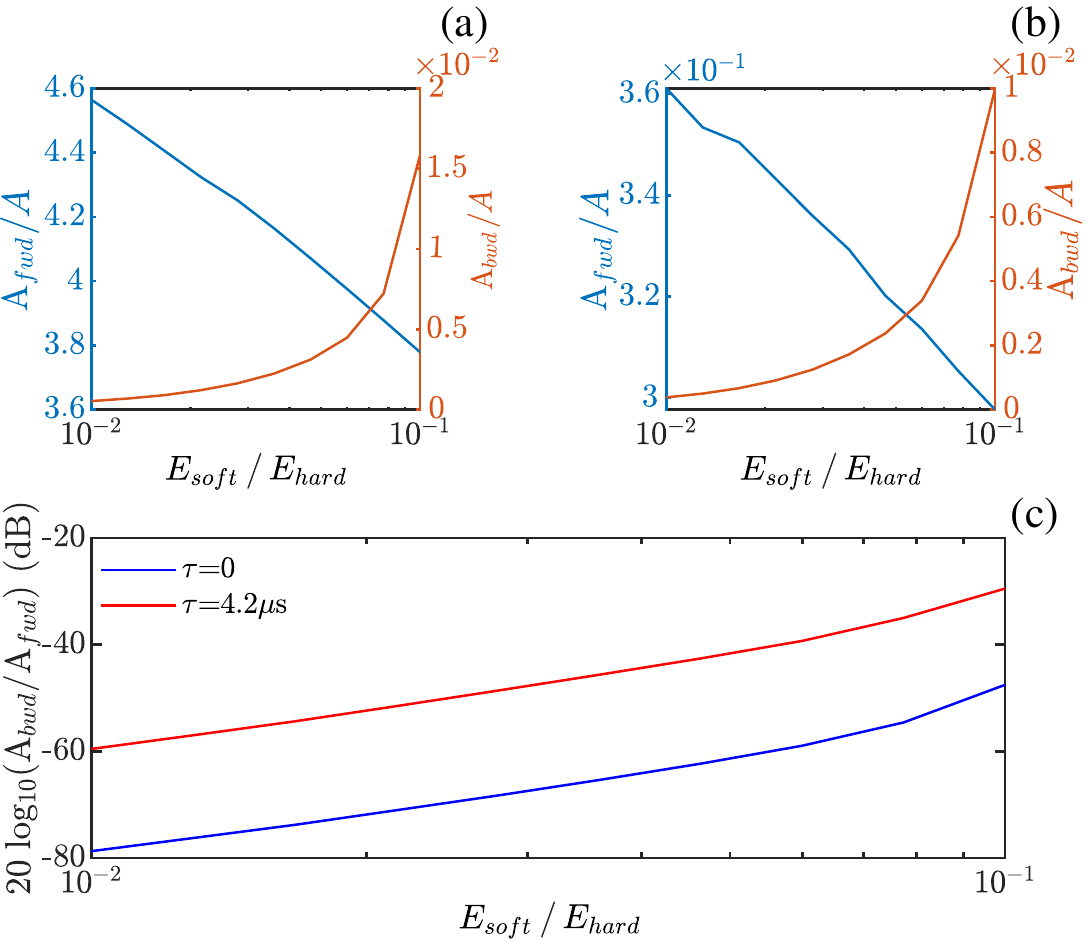}
\caption{Isolation level as a function of the elastic contrast $E_\mathrm{soft}/E_\mathrm{hard}$ at constant $E_\mathrm{hard}=E=4.5$~GPa and $\Omega=0.5$: displacement amplitudes of the LF demodulated signal in the forward ($A_\mathrm{fwd}$, $E_\mathrm{hard}\to E_\mathrm{soft}$) and in the backward ($A_\mathrm{bwd}$, $E_\mathrm{soft}\to E_\mathrm{hard}$) directions a) without ($\tau=0$) and b) with ($\tau=4.2$~$\mu$s) dissipation. c) Isolation ratio $A_\mathrm{bwd}/A_\mathrm{fwd}$ versus $E_\mathrm{soft}/E_\mathrm{hard}$ with and without dissipation.} 
\label{fig:11_diode_isolation}
\end{figure}

Figures~\ref{fig:10_diode_fwd_bwd}d and~\ref{fig:10_diode_fwd_bwd}j account for dissipative effects on the stepped diode. Contrasting these data with those of Figs.~\ref{fig:10_diode_fwd_bwd}b and~\ref{fig:10_diode_fwd_bwd}h in the time domain, \revision{and those of Figs.~\ref{fig:10_diode_fwd_bwd}c and~\ref{fig:10_diode_fwd_bwd}i in the frequency domain}, one observes that although the cutoff frequency is not precisely tuned for effective filtering, the dissipation rapidly attenuates the HF carrier wave \revision{(see the spurious frequency components close to the cutoff frequency around $600$~Hz in Fig.~\ref{fig:10_diode_fwd_bwd}c)}, facilitating the emergence of the LF demodulated signal. \revision{Also, the dissipation tends to compensate the frequency distortion due to the combined effect of dispersion and nonlinearity. Hence, the dissipation contributes to preserving the frequency content of the forward transmission, compare for instance the spectrum of the transmitted demodulated wave to that of the amplitude modulation in Fig.~\ref{fig:10_diode_fwd_bwd}e.} In the backward configuration, dissipation also ensures no transmission, except negligible leaks, when nonlinear effect are kept small. We observe that the losses significantly lower the amplitude of the LF demodulated signal in the forward direction, whereas it weakly affects the LF leaky component in the backward direction, see Figs.~\ref{fig:10_diode_fwd_bwd}d and~\ref{fig:10_diode_fwd_bwd}j.

Looking in details, the viscoelastic losses do not cancel the leaky transmission in the backward direction. The flaw resulting from a spurious demodulation in the soft chain, it can be resolved by making this chain more linear, e.g. by using even softer particles, see Fig.~\ref{fig:11_diode_isolation}. First, one observes that the magnitude of the forward demodulated signal is weakly affected (by no more than $20$\% over a decade) by the value of $E_\mathrm{soft}$, no matter dissipation is involved or not, see Fig.~\ref{fig:11_diode_isolation}a and~\ref{fig:11_diode_isolation}b. Indeed, in the range of probed elasticity, the soft chain efficiently filter the HF carrier since its cutoff frequency, $f_c^\prime=389$~Hz at $E_\mathrm{soft}/E_\mathrm{hard}=0.01$ and $f_c^\prime=838$~Hz at $E_\mathrm{soft}/E_\mathrm{hard}=0.1$, always remains lower than the HF carrier signal $f_d=903$~Hz. The leakage flaw thus does not rely on an imperfection of the filtering. In turn, decreasing $E_\mathrm{soft}$ increases the overlap $\delta_0$, linearizing further the wave propagation, and ultimately lowering the magnitude of the leaky LF component in the backward direction, see Fig.~\ref{fig:11_diode_isolation}a and~\ref{fig:11_diode_isolation}b. In practice, one achieves an isolation level close to $80$~dB with the softest particles probed, see Fig.~\ref{fig:11_diode_isolation}c. Non trivially, the dissipation lowers the isolation level by a constant $20$~dB in the whole range of probed elasticity, consistently with the previous observation that losses affect more the LF demodulated signal in the forward direction then in the backward direction, compare Figs.~\ref{fig:10_diode_fwd_bwd}(d,j) to Figs.~\ref{fig:10_diode_fwd_bwd}(b,h). 

\revision{Finally, it is worth emphasizing that the particles/ground interaction has been disregarded \-- whether it is elastic if their contact deforms or whether it is dissipative as in the case of solid friction or viscoelasticity \-- to focus on wave transmission in the axial direction only.  However, the grounding stiffness, equivalent to an onsite elastic reaction, would open a low frequency bandgap~\cite{Allein2020}, which would likely affect the generation and the propagation of the nonlinearly demodulated wave. In practice, the particles/ground interaction can be ruled out by using soft~\cite{Allein2020} or low friction lubricated support for instance, such that particle slide without without resistance or without rubbing. The design of an experimental prototype requires the consideration of a support; the analysis and understanding of the aforementioned ground effects are thus among the important objectives of a future study.}


\section{Conclusion}

In conclusion, we have shown that the features of the demodulated signal are strongly determined by three main effects: the nonlinearity, the wave dispersion and the energy dissipation. The effect of nonlinearity leads to the harmonic generation whose amplitude depends on the nonlinear parameter, $\epsilon$. For the form of the excitation selected in the present work, both the LF mode and the second harmonic are increasing function of $\epsilon$. \revision{Despite differences in the form of the chain excitation, wave packet in the present study,  the evolution of nonlinear modes are consistent with previous finding obtained using continuous wave as described by Tournat~\cite{Sanchez2013}.} 

\revision{The main system parameters are the dimensionless frequency and the dimensionless excitation amplitude. The former accounts for dispersion, while the latter controls the nonlinear effects.  Both parameters rule how fast the forward signal is demodulated, and their values at the diode junction can be controlled through the change on the elastic constants of chain segments.  It is shown that a ratio of elastic constants of 0.1 leads to backward leak of 0.01 whereas a ratio of 0.01 produces a negligeable leak without significantly affecting the forward direction.}

In the nonlinear regime and low dispersion ($\Omega<1/2$) the demodulated displacement of particles and the speed are proportional to $h(t)$ and $h'(t)$, respectively. By contrast, in the high dispersive regime ($\Omega>1/2$), the demodulated speed tends to the profile $h(t)$. This behavior is associated with the development of a step-compression front (LF mode) whose rise time is the characteristic modulation time of the carrier wave. In this case, the step front is followed by a flat plateau and the derivative on time of the front is proportional to $h(t)$. However, beads slowly relax to their equilibrium positions, which gives rise to a long tail whose derivative contribute to a slowly varying signal, as in Fig.~\ref{fig:08_mono_demodulation_LF_and_HF}(f,h).  The shape of demodulated wave reached after total mode conversion is consistent with analytical calculations carried out by Tournat et al.~\cite{Tournat2004}. 
 
The properties described above are crucial for a proper design of a nonlinear acoustic device such an acoustical diode. We showed that a stepped chain is suitable to implement such a device and the isolation level is close to 80 dB with a good choice of parameters.  Moreover, viscous dissipation in the low pass filter chain accelerates the elimination of the carrier wave improving the diode performance. In the backward configuration, if the carrier wave is selected above the cut-off frequency, the diode acts in the reverse configuration allowing no transmission. If the carrier frequency is below the cut-of frequency, dissipation quickly attenuates propagation, improving the rejection band of the diode. For a realistic choice of mechanical parameters, in the regime of low dispersion, in the forward configuration, only 20 hard beads provide sufficient growth of the self demodulated envelope, for which the beads speed is proportional to $h(t)$ and isolation level is close to 56 dB and 65 dB with and without losses respectively.


\section*{Acknowledgement}

F.M. acknowledges ANID-Chile through Fondecyt project N$^\circ$ 1201013 and Fondequip 130149. F.M. and S.J. acknowledges support from LIA-MSD France-Chile (Laboratoire International Associ\'e CNRS, ``Mati\`ere: Structure et Dynamique''). We acknowledge the support from DICYT Grant 042131MH-Postdoc of Universidad de Santiago de Chile. C.V. acknowledges ANID National Doctoral Program grant N$^\circ$ 21201036.


\section*{Statements}

\subsection*{Conflict of interest}
The authors declared that they have no conflicts of interest to this work. We do not have any commercial or associative interest that represents a conflict of interest in connection with the work submitted. The work described has not been submitted elsewhere for publication, in whole or in part, and all the authors listed have approved the manuscript that is enclosed.

\subsection*{Data availability}
The authors declare that the data supporting the findings of this study are available within the paper itself. Raw data and computer codes are available from the corresponding author on reasonable request.



\begin{thebibliography}{10}
\providecommand{\url}[1]{{#1}}
\providecommand{\urlprefix}{URL }
\providecommand{\doi}[1]{\url{https://doi.org/#1}}
\bibcommenthead

\bibitem{Engheta2006}
N.~Engheta, R.W. Ziolkowski, \emph{Metamaterials: physics and engineering
  explorations} (John Wiley \& Sons, 2006).
\newblock \doi{10.1002/0471784192}

\bibitem{Fok2008}
L.~Fok, M.~Ambati, X.~Zhang, Acoustic metamaterials.
\newblock MRS Bull. \textbf{33}(10), 931--934 (2008).
\newblock \doi{10.1557/mrs2008.202}

\bibitem{Guo2023}
X.~Guo, H.~Lissek, R.~Fleury, Observation of non-reciprocal harmonic conversion
  in real sounds.
\newblock Commun. Phys. \textbf{6}(1) (2023).
\newblock \doi{10.1038/s42005-023-01217-w}

\bibitem{Deymier2013}
P.A. Deymier (ed.), \emph{Acoustic Metamaterials and Phononic Crystals},
  \emph{Springer Series in Solid-State Sciences}, vol. 173 (Springer,
  Heidelberg, 2013).
\newblock \doi{10.1007/978-3-642-31232-8}

\bibitem{Smith2000}
D.R. Smith, W.J. Padilla, D.C. Vier, S.C. Nemat-Nasser, S.~Schultz, Composite
  medium with simultaneously negative permeability and permittivity.
\newblock Phys. Rev. Lett. \textbf{84}(18), 4184 (2000).
\newblock \doi{10.1103/PhysRevLett.84.4184}

\bibitem{Shelby2001}
R.A. Shelby, D.R. Smith, S.~Schultz, Experimental verification of a negative
  index of refraction.
\newblock Science \textbf{292}(5514), 77--79 (2001).
\newblock \doi{10.1126/science.1058847}

\bibitem{Smith2004}
D.R. Smith, J.B. Pendry, M.C.K. Wiltshire, Metamaterials and negative
  refractive index.
\newblock Science \textbf{305}(5685), 788--792 (2004).
\newblock \doi{10.1126/science.1096796}

\bibitem{Fleury2014}
R.~Fleury, D.L. Sounas, C.F. Sieck, M.R. Haberman, A.~Alù, Sound isolation and
  giant linear nonreciprocity in a compact acoustic circulator.
\newblock Science \textbf{343}, 516--519 (2014).
\newblock \doi{10.1126/science.1246957}

\bibitem{Wu2018}
Y.~Wu, M.~Yang, P.~Sheng, Perspective: Acoustic metamaterials in transition.
\newblock Journal of Applied Physics \textbf{123}(9), 090901 (2018).
\newblock \doi{10.1063/1.5007682}

\bibitem{Fang2020}
L.~Fang, A.~Darabi, A.~Mojahed, A.F. Vakakis, M.J. Leamy, Broadband
  non-reciprocity with robust signal integrity in a triangle-shaped nonlinear
  1d metamaterial.
\newblock Nonlinear Dyn. \textbf{100}(1), 1--13 (2020).
\newblock \doi{10.1007/s11071-020-05520-x}

\bibitem{Boechler2011}
N.~Boechler, G.~Theocharis, C.~Daraio, Bifurcation-based acoustic switching and
  rectification.
\newblock Nat. Mater. \textbf{10}(9), 665–--668 (2011).
\newblock \doi{10.1038/nmat3072}

\bibitem{Liang2010}
B.~Liang, X.S. Guo, J.~Tu, D.~Zhang, J.C. Cheng, An acoustic rectifier.
\newblock Nat. Mater. \textbf{9}, 989--992 (2010).
\newblock \doi{10.1038/nmat2881}

\bibitem{Popa2014}
B.I. Popa, S.A. Cummer, Non-reciprocal and highly nonlinear active acoustic
  metamaterials.
\newblock Nature Communications \textbf{5}(1), 3398 (2014).
\newblock \doi{10.1038/ncomms4398}

\bibitem{Li2011}
X.F. Li, X.~Ni, L.~Feng, M.H. Lu, C.~He, Y.F. Chen, Tunable unidirectional
  sound propagation through a sonic-crystal-based acoustic diode.
\newblock Phys. Rev. Lett. \textbf{106}, 084301 (2011).
\newblock \doi{10.1103/PhysRevLett.106.084301}

\bibitem{Maldovan2013}
M.~Maldovan, Sound and heat revolutions in phononics.
\newblock Nature \textbf{503}, 209–217 (2013).
\newblock \doi{10.1038/nature12608}

\bibitem{Devaux2015}
T.~Devaux, V.~Tournat, O.~Richoux, V.~Pagneux, Asymmetric acoustic propagation
  of wave packets via the self-demodulation effect.
\newblock Phys. Rev. Lett. \textbf{115}, 234301 (2015).
\newblock \doi{10.1103/PhysRevLett.115.234301}

\bibitem{Vakakis2009}
A.~Vakakis, O.~Gendelman, L.~Bergman, D.~McFarland, G.~Kerschen, Y.~Lee,
  \emph{Nonlinear Targeted Energy Transfer in Mechanical and Structural
  Systems} (Springer, New - York, 2009).
\newblock \doi{10.1007/978-1-4020-9130-8}

\bibitem{Vakakis2022}
A.F. Vakakis, O.V. Gendelman, L.A. Bergman, A.~Mojahed, M.~Gzal, Nonlinear
  targeted energy transfer: state of the art and new perspectives.
\newblock Nonlinear Dyn. \textbf{108}(2), 711--741 (2022).
\newblock \doi{10.1007/s11071-022-07216-w}

\bibitem{Grinberg2018}
I.~Grinberg, A.F. Vakakis, O.V. Gendelman, Acoustic diode: Wave non-reciprocity
  in nonlinearly coupled waveguides.
\newblock Wave Motion \textbf{83}, 49--66 (2018).
\newblock \doi{10.1016/j.wavemoti.2018.08.005}

\bibitem{Michaloliakos2023}
A.~Michaloliakos, C.~Wang, A.F. Vakakis, Machine learning extreme acoustic
  non-reciprocity in a linear waveguide with multiple nonlinear asymmetric
  gates.
\newblock Nonlinear Dyn. \textbf{111}(18), 17277--17297 (2023).
\newblock \doi{10.1007/s11071-023-08765-4}

\bibitem{Devaux2019}
T.~Devaux, A.~Cebrecos, O.~Richoux, V.~Pagneux, V.~Tournat, Acoustic radiation
  pressure for nonreciprocal transmission and switch effects.
\newblock Nat. Commun. \textbf{10}(1), 3292 (2019).
\newblock \doi{10.1038/s41467-019-11305-7}

\bibitem{Melo2006}
F.~Melo, S.~Job, F.~Santibanez, F.~Tapia, Experimental evidence of shock
  mitigation in a hertzian tapered chain.
\newblock Phys. Rev. E \textbf{73}(4), 041305 (2006).
\newblock \doi{10.1103/physreve.73.041305}

\bibitem{Job2007}
S.~Job, F.~Melo, A.~Sokolow, S.~Sen, Solitary wave trains in granular chains:
  experiments, theory and simulations.
\newblock Granul. Matter \textbf{10}(1), 13--20 (2007).
\newblock \doi{10.1007/s10035-007-0054-2}

\bibitem{Vasconcellos2022}
C.~Vasconcellos, R.~Zu\~niga, S.~Job, F.~Melo, Localized energy absorbers in
  hertzian chains.
\newblock Physical Review Applied \textbf{18}, 014078 (2022).
\newblock \doi{10.1103/PhysRevApplied.18.014078}

\bibitem{Job2005}
S.~Job, F.~Melo, A.~Sokolow, S.~Sen, How hertzian solitary waves interact with
  boundaries in a 1d granular medium.
\newblock Phys. Rev. Lett. \textbf{94}(17), 178002 (2005).
\newblock \doi{10.1103/PhysRevLett.94.178002}

\bibitem{Job2009}
S.~Job, F.~Santibanez, F.~Tapia, F.~Melo, Wave localization in strongly
  nonlinear hertzian chains with mass defect.
\newblock Phys. Rev. E \textbf{80}(2), 025602 (2009).
\newblock \doi{10.1103/physreve.80.025602}

\bibitem{Theocharis2010}
G.~Theocharis, N.~Boechler, P.G. Kevrekidis, S.~Job, M.A. Porter, C.~Daraio,
  Intrinsic energy localization through discrete gap breathers in
  one-dimensional diatomic granular crystals.
\newblock Phys. Rev. E \textbf{82}(5), 056604 (2010).
\newblock \doi{10.1103/physreve.82.056604}

\bibitem{Sanchez2013}
V.J. S{\'a}nchez-Morcillo, I.~P{\'e}rez-Arjona, V.~Romero-Garc{\'\i}a,
  V.~Tournat, V.E. Gusev, Second-harmonic generation for dispersive elastic
  waves in a discrete granular chain.
\newblock Phys. Rev. E \textbf{88}(4), 043203 (2013).
\newblock \doi{10.1103/PhysRevE.88.043203}

\bibitem{Kuwabara1987}
G.~Kuwabara, K.~Kono, Restitution coefficient in a collision between two
  spheres.
\newblock Jpn. J. Appl. Phys. \textbf{26}(8R), 1230 (1987).
\newblock \doi{10.1143/JJAP.26.1230}

\bibitem{Allein2020}
F.~Allein, V.~Tournat, V.~Gusev, G.~Theocharis, Linear and nonlinear elastic
  waves in magnetogranular chains.
\newblock Phys. Rev. Appl. \textbf{13}, 024023 (2020).
\newblock \doi{10.1103/PhysRevApplied.13.024023}

\bibitem{Tournat2004}
V.~Tournat, V.E. Gusev, B.~Castagnède, Self-demodulation of elastic waves in a
  one-dimensional granular chain.
\newblock Phys. Rev. E \textbf{70}(5), 056603 (2004).
\newblock \doi{10.1103/PhysRevE.70.056603}

\end{thebibliography}

\end{document}